\documentclass[useAMS,usenatbib]{article}

\usepackage{psfig}
\usepackage{epsfig}
\usepackage{graphicx}

\title{A parameter to quantify dynamics of a researcher's scientific
activity}
\author{S.B. Popov$^{1}$\thanks{E-mail:
polar@sai.msu.ru}\\
$^{1}$Sternberg Astronomical Institute,\\
 Universitetski pr. 13, 119992
Moscow, Russia}

\begin{document}

\maketitle

\begin{abstract}

I propose the  coefficient, $t_h$, and its modification $N_t$ which
in a simple way reflect dynamics of scientific activity of an individual 
researcher. I determine $t_h$ as a time period (from some moment in the past
till the present moment) during which papers responsible for 1/2 of the
total citation index were published. Parameter  $N_t$ represents average
of the citation index over this period:  $N_t={\mathrm{C.I.}}/2t_h$. 

\end{abstract}


\section{Introduction}

 The problem of estimation of an impact of a scientist (or a group of them)
is an actual one (see, for example,  \cite{redner04} and references therein). 
Still, in many countries, for example in
Russia, citation index (C.I. hereafter) or its modifications are not widely
used. Only now, especially in front of a possible reorganization in the sphere
of science, russian scientists and officials start to discuss 
problems related to
quantifying a scientific impact of individual researchers or their groups.

The task to quantify scientific output 
is non-trivial as many components are involved, and it is
impossible to describe fairly
quality of a scientist by a single parameter (to prove harmony by 
algebra\footnote{{\it "... dissected music like a
corpse, proved harmony by algebra..."} 
[A. Pushkin, transl. A. Shaw]}).
The total impact can be more or less given by the C.I. (we do not discuss
here such disadvantages of this parameter as dependence on research
topics, influence of promotion of results, personal contacts, etc.). 
However, the structure of
C.I. of a scientist (if it is mainly determined by a single paper with very
high C.I., or by several of them with medium C.I., or by numerous papers
with very small C.I., etc.) is lost when  only one simple parameter is used. 
Different modifications can be suggested. Recently, Hirsch \cite{hirsch05}
proposed an interesting coefficient which supplements the standard C.I. 
This parameter is sensitive to the structure of C.I., i.e. it can
describe if the index is dominated by few papers or not.
However, all these parameters do not reflect dynamics of scientific activity.
Below we propose a simple estimate which can distinguish if the C.I. of a
scientist is due to recent or old publications, so in principle it is
possible to estimate how it is probable that the scientist produce an
important result in near future.

\newpage

\section{Characteristic time}

 There were many attempts to include dynamics into bibliometric studies
(see, for example, the {\it citation age} in \cite{redner04} which reflects
the citation history of a given paper). 
It is important to determine a characteristic time interval not arbitrary,
but individually for each scientist. For example, one can think about a
minimum time (min $\Delta t$) in a scientific career of a person, 
when papers responsible for
1/2 of the total C.I. were published (obviously, for scientists who did the
main contribution in one paper or in a set of papers published during a
short time min $\Delta t$ is short, vise versa for those who continuosly
published papers of the same level min $\Delta t$ is comparable with the
duration of the career). If one adds to this min $\Delta t$ another
parameter - time interval separating the present moment $t_0$ from the
end of the period responsible for min $\Delta t$ - then we have
a rough figure of scientific activity of a scientists in time. However,
I think that a better parameter can be defined for clearness. 

Here I discuss a simple way to estimate a characteristic applicable to
individual scientists.  Up to my knowledge such a parameter was not
discussed before.
 
The idea is to define some characteristic time which can demonstrate how
long ago a scientist published papers which give the main contribution to
the C.I. I propose the parameter $t_h$ which is defined as follows.
It is the time (from the present moment towards the past) during which
papers that are responsible for 1/2 of the total C.I. were published.
 Let me examplify it.

\begin{figure}
\epsfxsize=0.72\hsize
\centerline{\epsfbox{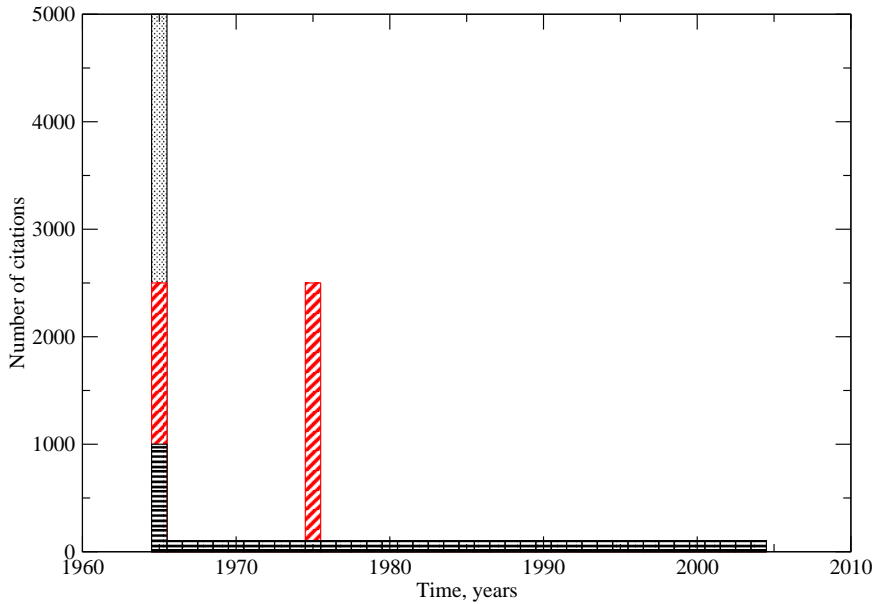}}
\caption{A simple illustration of scientific activity of three scientists
with the same total C.I., but with different distribution of important
papers over time. On the vertical axis I show the number of citations at the
present moment to papers published in a given year.}
\end{figure}

Imagine three scientists (see Fig.1). 
All started careers simultaneously. At the present
moment all three have the same C.I.=5000. One published in 1965 a paper with
C.I.=5000, and nothing after that. For him $t_h=40$~years. The second
published a paper with C.I.=2500 in 1965 and another one with the same C.I.
in 1975. For him we obtain $t_h=30$~years. The third one also had published in
1965 a top-cited paper with C.I.=1000, and then every year published a
papers all of which now have C.I.=100. For him $t_h=25$~years as 1/2 of his
C.I. is due to papers published after 1980.

All these values can be compared with another two limits (again we consider
scientists with career started in 1965, and with present-day C.I. equal to
5000). 
The first limit is the following, consider a
scientist with a constant rate of publications all papers of whom now have
equal number of citations would have $t_h=20$~yrs. Note, that
such a researcher actually demonstrate a growth of scientific output as
his/her later papers quicker gain citations.
Another limit (see also below) is a scientist with a constant rate of
publications all of which gain citations also with a constant rate. For
him/her we obtain $t_h\approx 28.3$~yrs. Clearly, among these five
scientists with equal C.I. those who demonstrated more activity recently has
shorter $t_h$.


Let us discuss the fifth case in more details (see Fig. 2.).
Consider a scientist who publishes papers with a constant rate during his
scientific career, and these papers recieve a constant number of citations
per year. In this case the number of citations of papers published in a
given year is proportional to time. Let us denote the coefficient (i.e. the
number of citations in a year per papers published in a given year) $A$ and
the durations of a career as  $T_{\mathrm{car}}$. 
So, we have $N_{\mathrm{cit}}=At$.  
The total number of citations is:

 $${\mathrm{C.I.}}=A T_{\mathrm{car}}^2/2$$.

We want to obtain $t_h$: 

$$
\frac12 {\mathrm{C.I.}}=A t_h^2/2=A  T_{\mathrm{car}}^2/4
$$

So, we obtain:

$$
t_h= T_{\mathrm{car}}/\sqrt 2.
$$

The value $T_{\mathrm{car}}/\sqrt 2$ is in some sense a critical one.
One can expect for scientists active in recent time
$t_h<T_{\mathrm{car}}/\sqrt 2$ independent on $T_{\mathrm{car}}$.

\begin{figure}
\epsfxsize=0.72\hsize
\centerline{\epsfbox{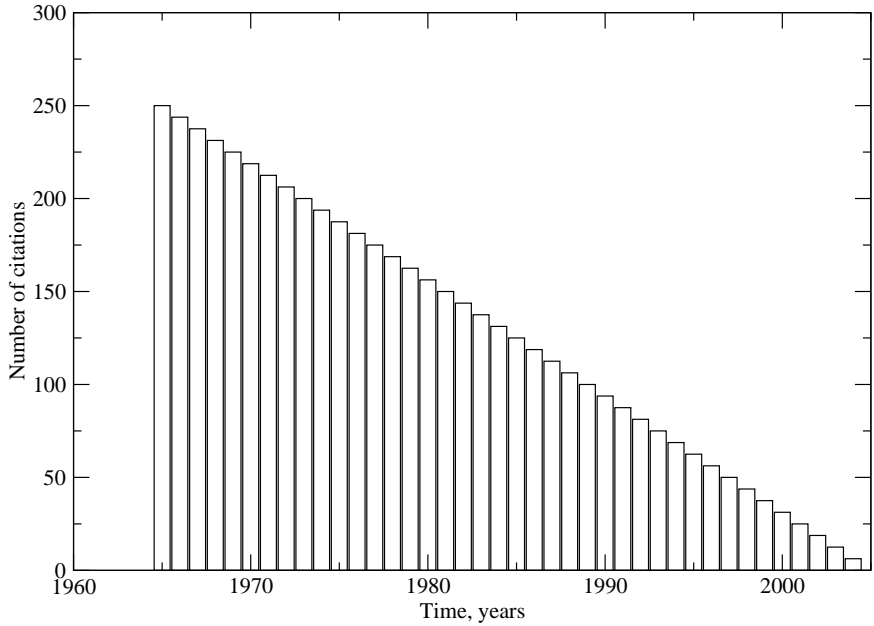}}
\caption{Constant citation rate per paper and contant rate of publications.
}
\end{figure}

 The parameter $t_h$ alone is not a very useful thing as it says nothing
about the total impact. But it can be useful to distinguish researchers
who's activity is not in the far past. 
Even for the same total C.I. $t_h$ is shorter for those who
published papers with large impact later.   
Especially, $t_h$ can be useful when both young and more senior scientists
are under consideration. It appears indeed capable to ideally complement the
standard C.I. or Hirsch's parameter $h$.

The main disadvantage of $t_h$ is the following: variations of this value
are not very large. However, as we propose this quantity as a secondary
coefficient (i.e. to compare researchers with similar C.I.) even small
differences (a factor $\sim 1.5$~--~2) are important.

 It is possible to modify $t_h$ to include information about the total C.I.
And in the following section I show a possible way to do it.

\newpage

\section{Average activity over the characteristic time}

 After we determine $t_h$ we know a characteristic time scale of scientific
activity of a researcher. Now what we can do is to average its C.I. (or
better 1/2 of C.I. as $t_h$ is related to half of the total index, and
letter {\it h} comes from {\it half}) over $t_h$.
We define 

$$
N_t={\mathrm{C.I.}}/2t_h.
$$

For the five scientists in the example above $N_t=$~62.5, 83.3, 100, 125, and
88.4 correspondently.
For the same total C.I. values of $N_t$ are different
demonstrating the fact that the first one was unactive for a long time, and
the periods of activity (and recognition of the results) 
of the third and the fourth 
ones are closer to the present moment than for
the rest two researchers (the fifth scientist in the example shows the same
rate of activity as the fourth, but rates of citation have different time
behaviour). For the same C.I.
$N_t$ can be different up to a factor of a  few 
(or even by an order of magnitude)
if persons have significantly different histories of scientific activity.

The main disadvantage of the approach is the possibility to have relatively
high $N_t$ just by self-citation, i.e. by publishing a huge amount of papers
refering to own previous publications. 


\section{Conclusions}  

I presented a simple estimate $t_h$ of a time scale which demonstrates
dynamics of scientific activity of scientists. Also the parameter
$N_t={\mathrm{C.I.}}/2t_h$  
was proposed as a compromise between integral and
dynamical characteristics of scientific impact. In my opinion $t_h$ can be a
good additional parameter to the standard C.I. value.

\section*{Acknowledgments}
 I want to thank Prof. J. Hirsch for comments on the idea of the coefficient
$t_h$, and participants of the projects Elementy.Ru, Scientific.Ru, and
astronomy.ru for discussions. The work was supported by the ``Dynasty''
Foundation (Russia).


\begin{thebibliography}{99}


\bibitem{redner04}Redner S., 2004, physics/0407137
\bibitem{hirsch05}Hirsch J.E., 2005, physics/0508025


\end{thebibliography}
\end{document}